\documentclass[conference, letterpaper]{IEEEtran}

\usepackage[boxruled,vlined,linesnumbered]{algorithm2e}

\ifCLASSINFOpdf
\else
\fi

\usepackage{subcaption}

\ifCLASSINFOpdf
   \usepackage[pdftex]{graphicx}
\else
\fi

\usepackage[cmex10]{amsmath}
\usepackage{color}
\usepackage{fancyhdr}

\renewcommand{\thispagestyle}[2]{}

\fancypagestyle{plain}{
        \fancyhead{}
        \fancyhead[C]{first page center header}
        \fancyfoot{}
        \fancyfoot[C]{first page center footer}
}
\begin{document}

%
\title{A New Approach to Speed up Combinatorial Search Strategies Using Stack and Hash Table}

\author{\IEEEauthorblockN{Bestoun S. Ahmed}
\IEEEauthorblockA{Istituto Dalle Molle di Studi \\sull’Intelligenza Artificiale (IDSIA)\\ 6928 Manno-Lugano, Switzerland\\ Salahaddin University - Erbil\\
Email: bestoun@idsia.ch}
\and
\IEEEauthorblockN{Luca M. Gambardella}
\IEEEauthorblockA{Istituto Dalle Molle di Studi \\sull’Intelligenza Artificiale (IDSIA)\\ 6928 Manno-Lugano, Switzerland\\
Email: luca@idsia.ch}
\and
\IEEEauthorblockN{Kamal Z. Zamli}
\IEEEauthorblockA{Faculty of Computer Systems\\ and Software Engineering\\
University Malaysia Pahang\\
Gambang, Malaysia\\
Email: kamalz@ump.edu.my}}

\maketitle

\begin{abstract}
Owing to the significance of combinatorial search strategies both for academia and industry, the introduction of new techniques is a fast growing research field these days. These strategies have really taken different forms ranging from simple to complex strategies in order to solve all forms of combinatorial problems. Nonetheless, despite the kind of problem these approaches solve, they are prone to heavy computation with the number of combinations and growing search space dimensions. This paper presents a new approach to speed up the generation and search processes using a combination of stack and hash table data structures. This approach could be put to practice for the combinatorial approaches to speed up the generation of combinations and search process in the search space. Furthermore, this new approach proved its performance in diverse stages better than other known strategies.
\end{abstract}


\IEEEpeerreviewmaketitle

\section{Introduction}
Combinatorial strategies have received lots of interest lately as a result of their diverse applications in areas of research, particularly in software engineering. In its simple form, a combinatorial strategy can reduce the several input parameters of a system to a small set of these parameters base on their interaction (combination) \cite{ref1}. This idea developed more recently to include the constraints and seeding among these input parameters also \cite{ref2, ref3}. Similarly, these parameters could be the features or configurations to be set for a system, or they might simply be the values to be entered while the system is in operation. The rationale behind the reduction is that it is impossible to take all possibilities of these input parameters. Therefore, the reduction must be done in a systematic way by considering the combinations of these parameters. As it is impossible to consider all likely combination for a system to test, there is a need to generate an optimised set of combinations that have the effectiveness of all possible combinations. 

Combinatorial strategies establish their effectiveness for different applications including software engineering, chemistry, biology, communication and many other fields \cite{ref4}. To this end, optimisation methods have been used to generate this set. Regardless of the optimisation technique used in the implemented strategies, evidence revealed that the serious issues in the development of these strategies are: how to construct the combinations and how to search for them later. To optimise, the strategy first has to generate all the possible combinations. Then, the optimisation algorithm attempts to cover these combinations with the smallest set. The complexity of this process is proportional with the number of input parameters. Hence, there is a need to speed up this process to enable the optimisation algorithms inside the combinatorial strategies work faster and efficiently. 

This paper proposes an approach to construct the combinations and also search for them efficiently. The approach includes new algorithm and programming techniques to construct, store and search for combinations. The remaining part of this paper is organised as follows: Section 2 gives a brief review of the combinatorial interaction strategies; Section 3 and 4 introduces the terminologies used all through this paper and similarly formulate the problem; Section 5 contains the methodology as well as the algorithms for the proposed strategy; Section 6 illustrates the experimental framework; Section 7 shows the experimental results and finally, Section 8 presents our conclusions.

\section{Combinatorial Interaction Strategies}

In the last decade, various studies on combinatorial interaction approaches have gained a lot of awareness in such a way that several test generation approaches were developed.  With the approaches generally dedicated to solving different problems, a few of them solve the generation of optimised set of input parameters by taking combinations into account. Though some others are also dedicated to generating those sets with constraints or seeding among the input factors, they still require particular configuration. Other research groups have started to examine (instead of software engineering alone) the application of these approaches in other research fields like biology, chemistry and electrical engineering to solve real life problems.

Evidence revealed that the use of meta-heuristic algorithms could achieve optimum sets of final combinatorial set covering every interaction among the input parameters. Most recently, different meta-heuristic algorithms have been adapted to solve this problem such as Simulated Annealing (SA) \cite{ref5}, Genetic Algorithms (GA) \cite{ref6}, Cuckoo Search (CS) \cite{ref7} and many other algorithms. Despite the wide range of approaches and algorithms used in generating the combinatorial interaction set, we still cannot find a “universal” strategy that can generate optimised sets for all the configuration since this problem is an NP-complete problem \cite{ref8}. Hence, each strategy could be useful for specific kind of configuration and application.

Although different strategies have been developed, the problem of search space complexity is still the same. As mentioned earlier, the main aim of the combinatorial strategies is to cover the entire interactions of input parameters by the smallest set. Hence, the strategy needs to search for a combination that can cover much of those interactions. To determine the number of interactions covered, the strategy must search for them among a large number of interactions which will definitely consume the program time as well as the resources of the computer. It will likewise cause the program to take more iteration for searching within the meta-heuristic algorithm. 

In addition to the aforementioned issue, the problem of generating input parameter combination represents another serious problem apart from consuming time and resources. This problem appears worsen as the input parameter continues to grow in size since most of the algorithms’ complexities are growing with the number of parameters. To overcome this problem, a special algorithm is needed to be combined with efficient data structures in order to speed up the generation and sorting process. This paper aims to provide new approaches and algorithms that will solve these problems and at the same time, speed up the combinatorial interaction search strategies in general.

\section{Preliminaries}
Combinatorial interaction strategies relies on Covering Array (CA) a well-known mathematical model to represent the combinatorial interaction set. The CA notation assures that all the interactions represented within one array. This mathematical object originates essentially from another object called orthogonal array (OA) \cite{ref9}. An $OA_\lambda (N; t, k, v)$ is an $N \times k$ array, where for every $N \times t$ sub-array, each $t-tuple$ occurs exactly $\lambda$ times, where $\lambda = N/v^t$; $t$ is the combination strength; $k$ is the number of input factors ($k \geq t$); and $v$ is the number of symbols or levels associated with each input factor. In covering all the combinations, each $t-tuple$ must occur at least once in the final test suite \cite{ref10}. When each $t-tuple$ occurs exactly once, $\lambda=1$, and it can be unmentioned in the mathematical syntax, that is, $OA (N; t, k, v)$. As an example, $OA( 9; 2, 4, 3)$ contains three levels of value ($v$) with a combination degree ($t$) equal to two, and four factors ($k$) can be generated by nine rows.

CA is another mathematical notation that is more flexible in representing test suites with larger sizes of different parameters and values. In general, CA uses the mathematical expression $CA_\lambda (N; t, k, v)$ \cite{ref11}. A $CA_\lambda(N ; t, k, v)$ is an $N \times k$ array over $(0, . . . , v-1)$ such that every $B={b_0, ..., b_{t-1}}$ $\ni $ is $\lambda$-covered and every $N\times t$ sub-array contains all ordered subsets from $v$ values of size d at least $\lambda$ times \cite{ref12}, where the set of column $B=b_0, ..., b_{t-1} \supseteq {0, ..., k-1}$.To ensure optimality, we normally want $t-tuples$ to occur at least once. Thus, we consider the value of $\epsilon=1$, which is often omitted. The notation becomes $CA(N;t,k,v)$ \cite{ref13}. Based on this notation and since the strategy is mainly depending on the interaction degree ($t$), the combinatorial strategies are sometimes termed $t-way$ strategies.

\section{Generation of n-combinations}

Different algorithms have been employed in the literature to generate the combinations of input parameters. The most common among them all is the n-bit enumerator. As the name implies, the algorithm starts by enumerating from $0$ to $2^{n – 1}$ ($n$ is the number of input parameters) thereafter, it filters the number base on the specified combination strength. For example, when $t = 4$, the algorithm will only choose those numbers with four true cases and neglect the rest. $N-bit$ enumerator has been used in different combinatorial search strategies in different ways as well as in other research fields  (ex. \cite{ref14}). Often, working perfectly when the number of input parameters is small, however, owing to the complexity of the algorithm $O(n)$, $n-bit$ enumerator becomes complex and the generation process tends to get slower as the number of input parameters increases.

Nowadays, software systems are prone to complexity in different ways due to many configurations and feature that may be present in any software to ensure its functions are properly managed. For example, Software Product Lines (SPL) need several parameters to be adjusted with different configurations due to different products that must be tested. To this end, other algorithms have also been implemented in the literature to overcome this issue. So far, backtracking, counting, and subsets algorithms have been used in the literature to solve this problem by speeding up the construction of t-combinations. Though each of the algorithms has its own approach to constructing the combinations, they also have different limitations for the input parameters and performance. Hence, this paper proposes another approach not only to speed up the construction but to search for them in an efficient way.

\section{Methodology}
\subsection{The Generation of Parameter Combination }

The algorithm used the CA notation as the base of input. As shown in Algorithm 1, the algorithm took $k$ input parameters and produced t-combination of them each time adding the combinations to a final array containing all $t$-combinations of $k$. Instead of enumerating all $n-bits$, a stack data structure was used to hold the parameters permanently by “push” them into the stack and then “pop” them when needed during the iterations. Additionally, a temporary array was created with index i to help the generated combinations in each iteration (Steps 1-2). A stack data structure ($S$) was created and the first parameter (0) was pushed inside (Steps 3-4). The algorithm continued to iterate until the stack became empty (Step 5). The index number $i$ of the Comb array was set to length of $S -1$ and the value $v$ of this index $i$ was set to the top value in the stack (i.e. pop) until the $v$ was less than $k$ (Steps 6-9). Furthermore, the algorithm continued to increment the $i$ and $v$ then pus the value of $v$ into $S$ until the index number was equal to the length of the required interaction strength t (Steps 9-15). The pseudo code is shown clearly in Algorithm 1. Additionally, for better understanding of the algorithm, a running example is illustrated in Figure 1.\\



\begin{algorithm}
 \KwIn{Input-parameters $k$ and combination strength $t$}
 \KwOut{All $t$-combinations of k where $k={k_1 , k_2 , k_3 , â€¦ k_n}$ }
 Let Comb be an array of length $t$\;
 Let $i$ be the index of Comb array\;
 Create a stack $S$\;
 $S \leftarrow 0$\;
 \While {$S\neq null$}{
 
$i$ =(the length of $S -1$)\;
$v$ = pop the stack value\;

\While {pop value $< k$}{
set Comb of index ($i$) to $v$\;
$i\leftarrow i + 1$\;
$ v\leftarrow v + 1$\;
push $v$ to stack\;

\If{$i=t$}{
Add Comb to final array\;
break\;
}

}
}
 \caption{Parameter Combination Generator}
\end{algorithm}

Figure 1 shows a running example to illustrate how the combinations of input parameters were generated using three input parameters [0, 1, and 2]. With the first parameter pushed into the stack at start, the algorithm iterated and the stack pop its last value to the $i+1$ index of the Comb array. In the next iteration, the stack was pushed by $v+1$ value. The algorithm stopped when the stack became empty. The final array then contained all the interaction of input parameters which are, [(0:1), (0:2), (1:2)].

\begin{figure} [h!]
\centering

\includegraphics [width=\linewidth]{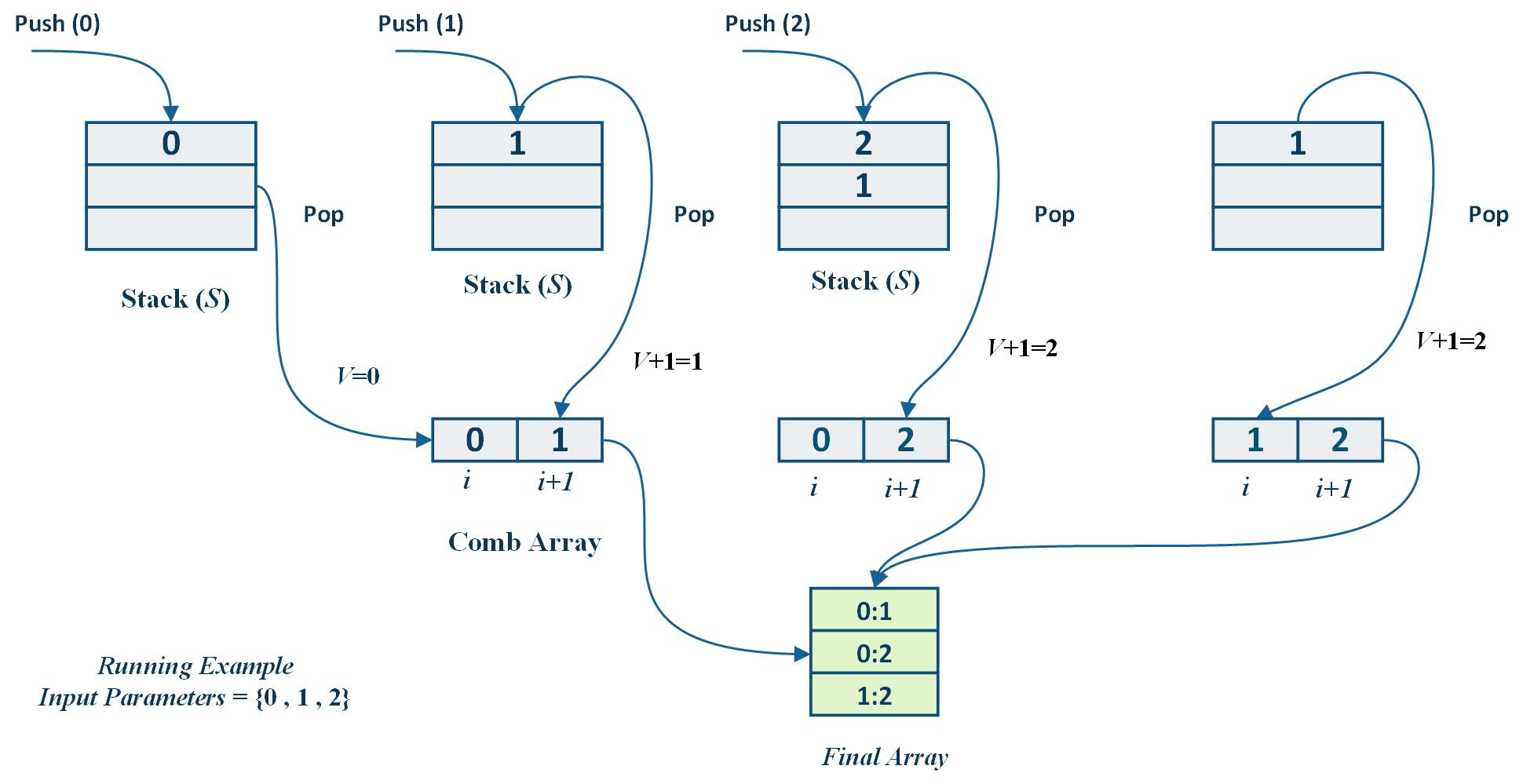}

\caption{A running example for the algorithm in Algorithm 1}

\end{figure}

As could be seen in Figure 1, the algorithm kept the previous value of $v$ for the next iteration unless it became greater than the t value. For example, $v = 0$ in the first iteration and in the next iteration, it became $v+1$ which equals to 1. Then it was incremented and pushed into the stack again.

\subsection{Searching for the Interaction Coverage}

The combinatorial search strategies need to generate all the possible interaction elements between the input parameters. This step is vital so as to verify how many of these elements can be covered by the suggested solution. Most of the time, this will be the fitness function of the meta-heuristic used in the strategy. It is not clear in most of the implemented strategies which data structure and searching mechanism they used since they are close sourced. However, for the known strategies, there are different mechanisms to store and search for the interaction elements.

The elements could be saved in a database and search later. The searching process could be enhanced by using a kind of indexing mechanism when storing them. However, these will potentially slowdown the search as there could be another outside system that may need to be interfaced with. Thus, another direction is to store the elements in the same program in an array and then search for them.

\begin{figure} [h!]
\centering
\includegraphics [width=\linewidth]{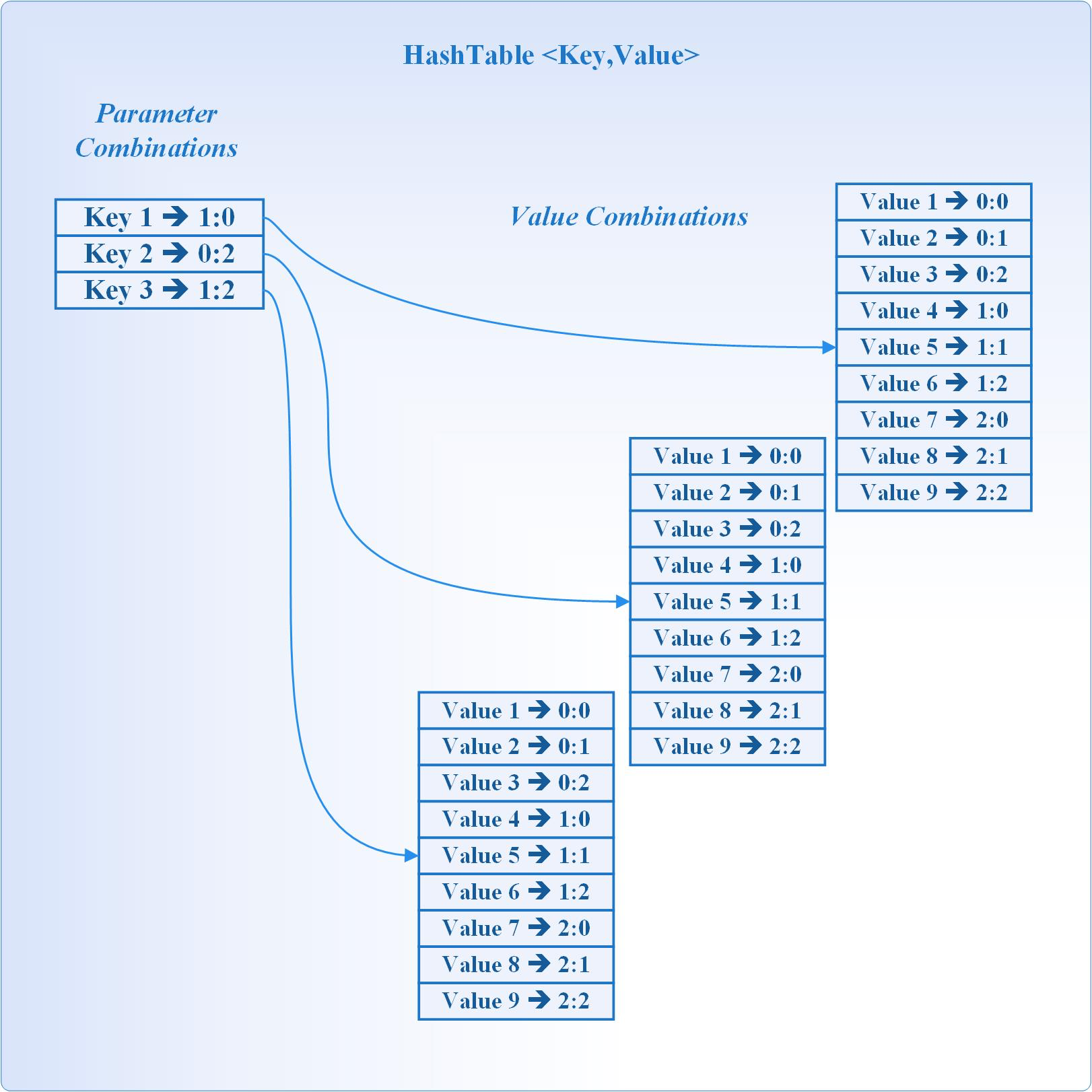}
\caption{Representation of the adopted HashTable Structure}
\end{figure}

Since there will be a huge amount of elements, the time for searching will increase dramatically with the increase of the parameter and value numbers. To overcome this issue, an indexing mechanism was used to store the elements base on the interaction in a sorting array and then search for a specific element in its corresponding combination. This could speed up the search processes effectively but, the time required for finding a specific element will increase when the number of values increases (since there will be several elements with equal parameter combination). Hence, there is a need to find a new approach to store and search for the interaction elements efficiently.

The proposed approach in this paper uses the HashTable data structure to store the interaction elements. As shown in Figure 2, the data structure is composed of <Key, Value> pairs, and the elements are stored based on the key with each key holding different value. When the program needs to know the number of interaction elements that could be covered by the possible solution, it will send it to the search function. The search function then searches for the interaction element in the exact <key, value> pair. Hence, without the need to search the entire element set, the function knows the location of the specified element.

\section{Experimental Framework}
To evaluate the proposed approach, two sets of experiments were performed. The first set of experiment was to evaluate the performances of the parameter combination generation algorithm. This was carried out by running the algorithm under different conditions of input parameters and interaction degrees. In addition, other algorithms were implemented within the same environment to compare the results with them.

In the second set of experiment, the performances of the search mechanism were evaluated and different sets of benchmark were considered. The benchmarks are varied in the number of input parameters and in individual value. The performance is defined by the time it takes the algorithm to find the set of interaction elements for a specific solution. For the purpose of comparison, two other mechanisms were used in the experiments. The first mechanism stored all interaction elements in an array and then search for the elements while the second mechanism, stored all elements in an indexed array and then search for the elements.

All experiments are conducted within an environment of desktop computer with windows 10 installed, CPU 2.9 GHz Intel Core i5, 8 GB 1867 MHz DDR3 RAM, 512 MB of flash HDD. The algorithms are implemented in .Net environment.

\section{Experimental Results and Discussion}

As mentioned earlier, the experiments were performed within two phases. The parameter combination generation algorithm with different parameter size interaction strength was evaluated. The parameters were varied from 20 to 400 input parameters. It is worth to mention here that the algorithm can take more than 1000 parameters as input. However, there was no evidence in the literature showing the use of more than this amount of parameter. In addition, the interaction strength was varied from 2 to 6 since this was the range of interactions used in the research so far. Figure 3 shows the comparison of these results, with the x-axis showing the parameter sizes and the y-axis showing the time in milliseconds in logarithmic scale.  

The results showed several important points about the algorithm, and it could be noted that the algorithm performed very well for the generation. Also, it could generate the combination of 400 parameters when $t = 2$ with less than 5 milliseconds. The performance dropped when the interaction strength became higher as could be seen in the figure. However, it still performed well. For example, it could generate the combination of 100 parameters when $t = 6$ with less than 60 seconds. The drop in performance was due to the stack capacity and the several parameters pushed into the stack as the interaction strength increases. It could also be noted from the algorithm that when the interaction strength becomes higher, for example ($t = 6$), 6 parameters should be push and pop each time. This will slow down the algorithm.

\begin{figure} [h!]
\centering
\includegraphics [width=\linewidth]{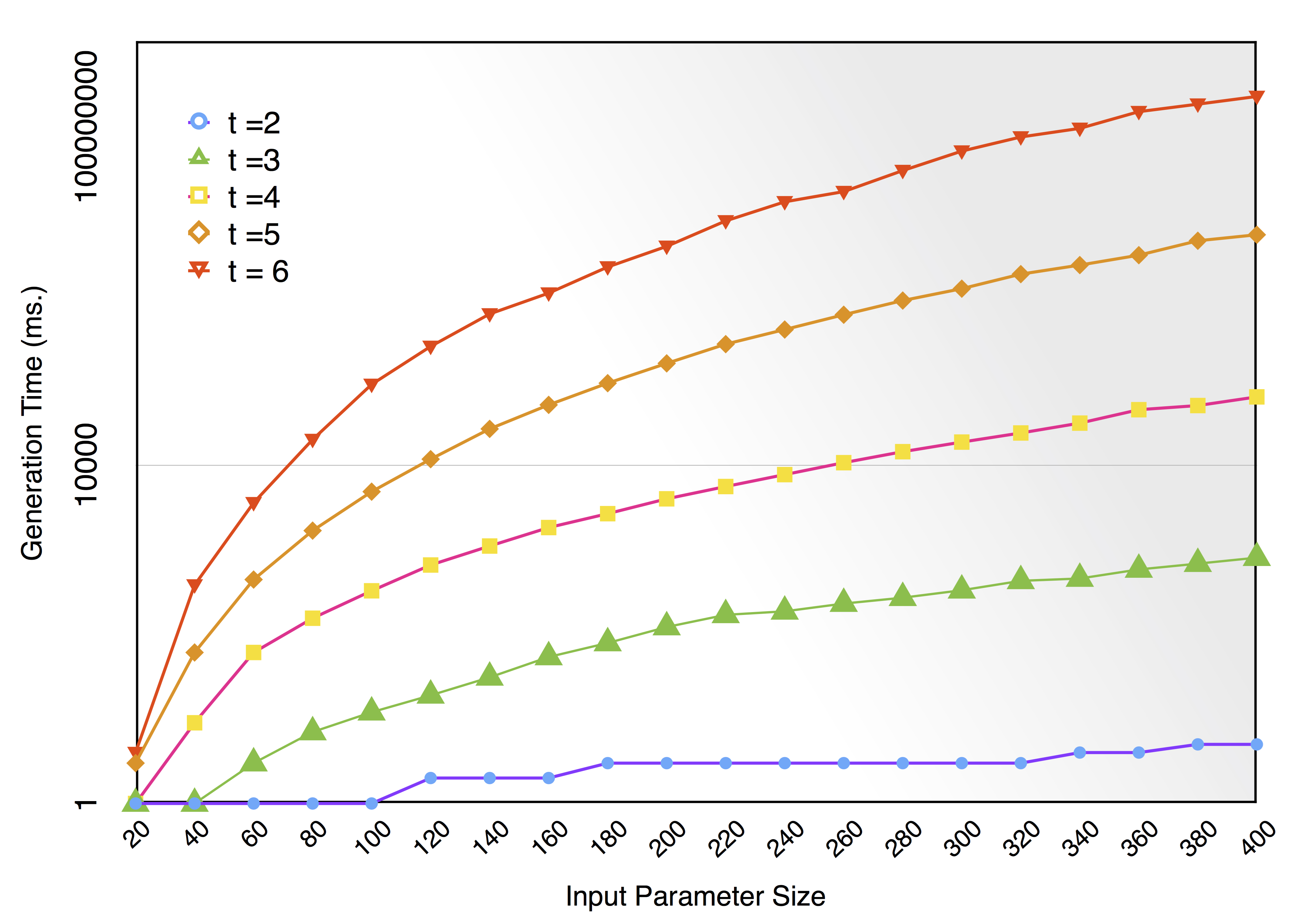}
\caption{Performance of the algorithm with the variation of input parameter and interaction strength}
\end{figure}

The second set of experiment was conducted to compare the algorithm with the existing available algorithms. The n-bit enumerator is implemented within our environment for comparison since it is the fundamental algorithm in this direction. When the algorithm was executed, we observed that the algorithm could not generate combinations more than 30 parameters within our environment due to enough memory exception. The limitation of the algorithm was due to the compiler and memory limitations since they could not perform large variable when the parameter ‘values became higher. However, it should be mentioned that the algorithm takes 546 ms for generating combinations of 20 parameters.

The last set of experiment was the search time in the search space. The search time for the relevant interaction elements was measured. This time indicated the maximum time taken by the algorithm to discover the relevant interaction elements for a specific solution. The maximum time was taken because the time may vary and decrease as the algorithm iterates since some of the interaction elements will be deleted. Hence, the maximum time gave a good indication about the time taken by the algorithm when the search space was full. Figures 5 and 6 show this time when $t = 2$ and 3 respectively for two different benchmarks.

As could be noted from the figures, two configurations were taken in the experiments for a covering array generation. The configurations were $CA (N; 2, 10 10)$ and $CA (N; 2, 10 20)$ in which the interaction strength $t = 2$ then $CA (N; 3, 10 10)$ and $CA (N; 2, 10 20)$ where the interaction strength $t = 3$. The configurations represent perfect benchmarks for this experiment since they have many parameters and many values for of the parameters. This will make the search space more complicated with many interaction elements.

\begin{figure}
\centering
\includegraphics [width=\linewidth]{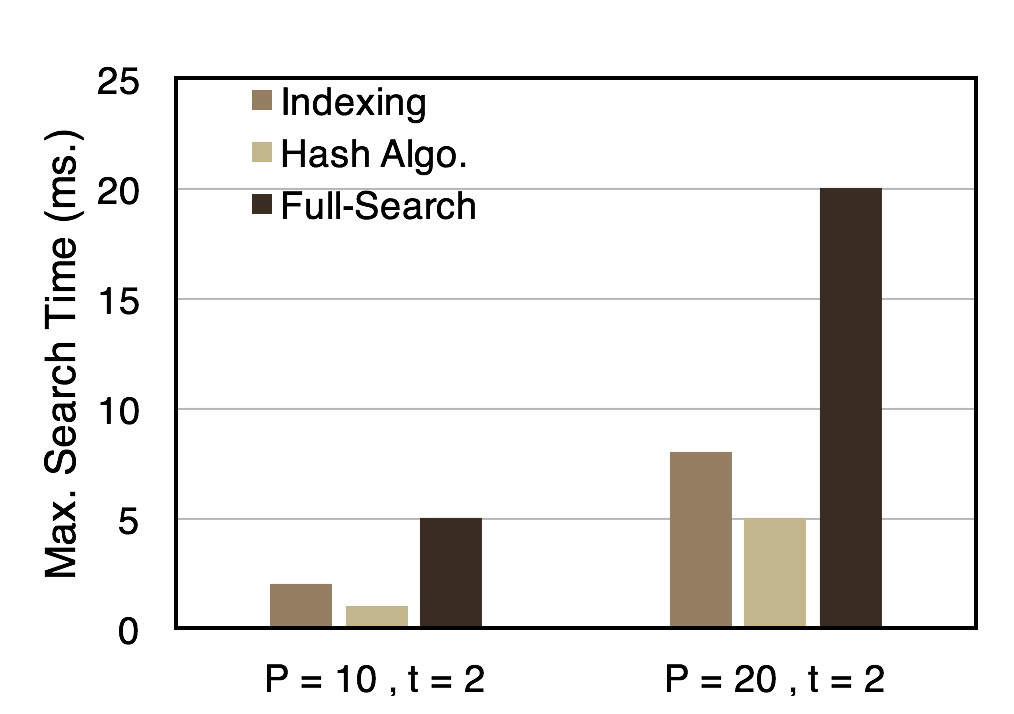}
\caption{Maximum search time measured for $t=2$ when $v=10$ and $P = 10$ and 20 respectively}

\end{figure}

\begin{figure}
\centering
\includegraphics [width=\linewidth]{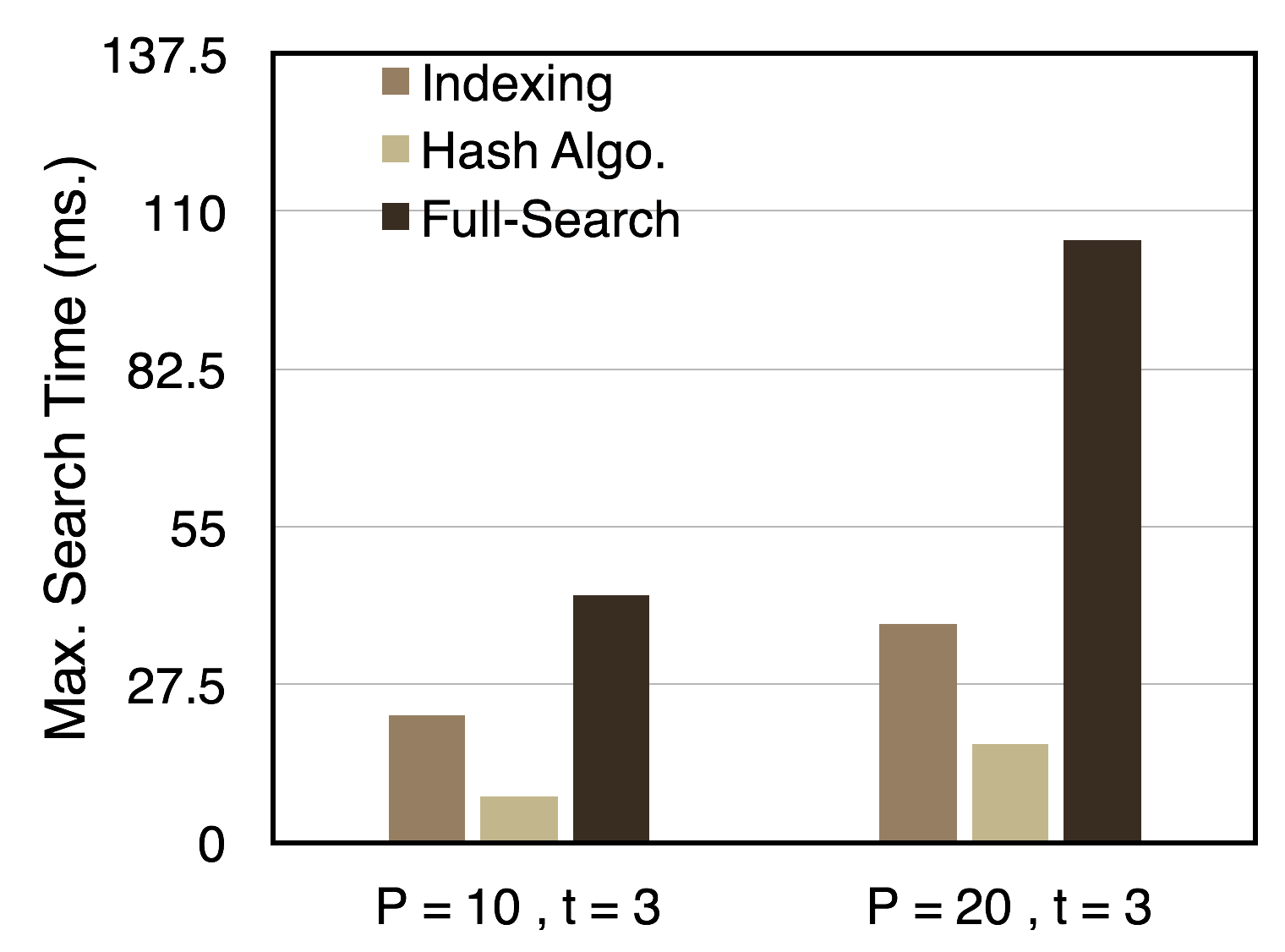}
\caption{Maximum search time measured for $t=3$ when $v=10$ and $P = 10$ and 20 respectively}

\end{figure}

As could be noticed from Figures 4 and 5, the maximum search time for the interaction elements was compared for three searching mechanism, “Hash Algo”, which is our mechanism with “Full-Search” and “indexing” mechanism. As mentioned earlier, the indexing mechanism saved the interaction elements in a sorted array and stored the indexing of each group of elements, however, “Full-Search” mechanism sort all the elements in an array without indexing and hence search exhaustively every time in Full-Search for all elements.

Both Figure 4 and 5 showed that our mechanism reduced the search time dramatically and hence can improve the total generation time of the solution. The figures also show that the “Full-Search” mechanism took more time to find the related interaction elements for a solution. The indexing mechanism showed better performance as compared to “Full-Search.” The searching time for the indexing mechanism was low when the number of parameters and values were low. However, as they were getting higher or the interaction strength is getting higher, the performance dropped due to the many interaction elements that must be searched for in one group of indexing. Our mechanism performed better in a dramatic way as compared to other mechanisms.

For example in Figure 4, when P=10 and t=2, the search time for our mechanism was less than 1(ms), while the indexing took more than 2 (ms). When P=20, our mechanism took less than 5(ms) for search while indexing took more than 8(ms). This improvement in performance could be seen clearly in case of t=3. When P=10, our mechanism took less than 8 (ms) for search and indexing took more than 22 (ms) whereas the “Full-Search” took more than 43 (ms).

The performance of other mechanisms continues to drop in case of t=3 when P=20 in which the indexing search time became 38 (ms) and our mechanism was 17 (ms). It should be mentioned here that this performance of the search affected the totally performance cumulatively since this search process is the most consuming computation in the combinatorial search strategies. 

\section{Conclusion}
In this paper, we have presented our proposed approach to generate and search for the interaction elements of the input parameters of the combinatorial search strategies. Based on our experience with these strategies, the generation of input parameters’ combinations and search for the interaction elements for the fitness function will slow down the generation process of the final test suite of the interaction. This paper serves as a guide and framework for future implementation of combinatorial strategies. The implemented approach proved its performance for generation input parameters faster than other algorithms for difference sizes and also its performance is faster than other algorithms when searching for the interaction elements. This paper is part of an existing research on combinatorial interaction testing for generating effective test cases for different applications.
\section*{Acknowledgment}

The first author of the paper would like to thanks IDSIA institute and Swiss Excellence Scholarship for hosting and supporting this research.

\bibliography{references}

\begin{thebibliography}{10}
\providecommand{\url}[1]{#1}
\csname url@samestyle\endcsname
\providecommand{\newblock}{\relax}
\providecommand{\bibinfo}[2]{#2}
\providecommand{\BIBentrySTDinterwordspacing}{\spaceskip=0pt\relax}
\providecommand{\BIBentryALTinterwordstretchfactor}{4}
\providecommand{\BIBentryALTinterwordspacing}{\spaceskip=\fontdimen2\font plus
\BIBentryALTinterwordstretchfactor\fontdimen3\font minus
  \fontdimen4\font\relax}
\providecommand{\BIBforeignlanguage}[2]{{%
\expandafter\ifx\csname l@#1\endcsname\relax
\typeout{** WARNING: IEEEtran.bst: No hyphenation pattern has been}%
\typeout{** loaded for the language `#1'. Using the pattern for}%
\typeout{** the default language instead.}%
\else
\language=\csname l@#1\endcsname
\fi
#2}}
\providecommand{\BIBdecl}{\relax}
\BIBdecl

\bibitem{ref1}
C.~Nie, H.~Wu, X.~Niu, F.-C. Kuo, H.~Leung, and C.~J. Colbourn, ``Combinatorial
  testing, random testing, and adaptive random testing for detecting
  interaction triggered failures,'' \emph{Information and Software Technology},
  vol.~62, no.~0, pp. 198--213, 2015.

\bibitem{ref2}
R.~C. Bryce and C.~J. Colbourn, ``Prioritized interaction testing for pair-wise
  coverage with seeding and constraints,'' \emph{Information and Software
  Technology}, vol.~48, no.~10, pp. 960--970, 2006.

\bibitem{ref3}
L.~Yu, F.~Duan, Y.~Lei, R.~N. Kacker, and D.~R. Kuhn, ``Constraint handling in
  combinatorial test generation using forbidden tuples,'' in \emph{Software
  Testing, Verification and Validation Workshops (ICSTW), 2015 IEEE Eighth
  International Conference on}, 2015, Conference Proceedings, pp. 1--9.

\bibitem{ref4}
B.~S. Ahmed and K.~Z. Zamli, ``A review of covering arrays and their
  application to software testing,'' \emph{Journal of Computer Science},
  vol.~7, no.~9, pp. 1375--1385, 2011.

\bibitem{ref5}
B.~S. Ahmed, K.~Z. Zamli, and C.~P. Lim, ``Application of particle swarm
  optimization to uniform and variable strength covering array construction,''
  \emph{Applied Soft Computing}, vol.~12, no.~4, p. 1330–1347, 2012.

\bibitem{ref6}
T.~Shiba, T.~Tsuchiya, and T.~Kikuno, ``Using artificial life techniques to
  generate test cases for combinatorial testing,'' in \emph{28th Annual
  International Computer Software and Applications Conference}.\hskip 1em plus
  0.5em minus 0.4em\relax IEEE Computer Society, 2006, Conference Proceedings,
  pp. 72--77 vol.1.

\bibitem{ref7}
B.~S. Ahmed, T.~S. Abdulsamad, and M.~Y. Potrus, ``Achievement of minimized
  combinatorial test suite for configuration-aware software functional testing
  using the cuckoo search algorithm,'' \emph{Information and Software
  Technology}, vol.~66, no.~0, pp. 13--29, 2015.

\bibitem{ref8}
K.~C. Tai and Y.~Lie, ``In-parameter-order: a test generation strategy for
  pairwise testing,'' in \emph{3rd IEEE International Symposium on
  High-Assurance Systems Engineering}.\hskip 1em plus 0.5em minus 0.4em\relax
  IEEE Computer Society, 1998, Conference Proceedings, pp. 254--261.

\bibitem{ref9}
C.~S. Cheng, ``Orthogonal arrays with variable numbers of symbols,'' \emph{The
  Annals of Statistics}, vol.~8, no.~2, pp. 447--453, 1980.

\bibitem{ref10}
A.~H. Ronneseth and C.~J. Colbourn, ``Merging covering arrays and compressing
  multiple sequence alignments,'' \emph{Discrete Applied Mathematics}, vol.
  157, no.~9, pp. 2177--2190, 2009, 1537488.

\bibitem{ref11}
R.~N. Kacker, D.~Richard~Kuhn, Y.~Lei, and J.~F. Lawrence, ``Combinatorial
  testing for software: An adaptation of design of experiments,''
  \emph{Measurement}, vol.~46, no.~9, pp. 3745--3752, 2013.

\bibitem{ref12}
C.~Colbourn, G.~Kéri, P.~R. Soriano, and J.-C. Schlage-Puchta, ``Covering and
  radius-covering arrays: Constructions and classification,'' \emph{Discrete
  Applied Mathematics}, vol. 158, no.~11, pp. 1158--1180, 2010.

\bibitem{ref13}
A.~Hartman and L.~Raskin, ``Problems and algorithms for covering arrays,''
  \emph{Discrete Mathematics}, vol. 284, no. 1-3, pp. 149--156, 2004.

\bibitem{ref14}
N.~Alon, O.~Goldreich, J.~Håstad, and R.~Peralta, ``Simple constructions of
  almost k-wise independent random variables,'' \emph{Random Structures and
  Algorithms}, vol.~3, no.~3, pp. 289--304, 1992.

\end{thebibliography}
\bibliographystyle{IEEEtran}

\end{document}